\documentclass[aps, prb, reprint, twocolumn, superscriptaddress]{revtex4-1}
\usepackage{enumitem}
\usepackage{multirow}
\usepackage{graphicx}
\usepackage{longtable}
\usepackage[utf8]{inputenc}
\usepackage[T1]{fontenc}
\usepackage{epstopdf}
\usepackage{textcomp}
\usepackage[cmyk,dvipsnames]{xcolor} % For obvious reasons
\usepackage{amsbsy}
\usepackage{amsmath}
\usepackage{amssymb}
\usepackage{amsfonts}
\usepackage{dsfont}
\usepackage{mathtools}
\usepackage{braket}
\usepackage{array}
\usepackage{gensymb}
\usepackage{placeins}
\usepackage{dashrule}
\usepackage{xcolor}
\usepackage{booktabs}
\usepackage{hyperref}
\usepackage{float}
\usepackage{braket}
\usepackage{ulem}

\makeatletter
  \def\my@tag@font{\normalsize}
  \def\maketag@@@#1{\hbox{\m@th\normalfont\my@tag@font#1}}
  \let\amsmath@eqref\eqref
  \renewcommand\eqref[1]{{\let\my@tag@font\relax\amsmath@eqref{#1}}}
\makeatother

\allowdisplaybreaks

\begin{document}

\title{Control of stripe domain wall magnetization in perpendicular anisotropy multilayers}

\author{Ruslan Salikhov}
\email{r.salikhov@hzdr.de}
\affiliation{Institute of Ion Beam Physics and Materials Research, Helmholtz-Zentrum Dresden-Rossendorf, Bautzner Landstrasse 400, 01328 Dresden, Germany}

\author{Fabian Samad}
\affiliation{Institute of Ion Beam Physics and Materials Research, Helmholtz-Zentrum Dresden-Rossendorf, Bautzner Landstrasse 400, 01328 Dresden, Germany}
\affiliation{Institute of Physics, Chemnitz University of Technology, Reichenhainer Strasse 70, 09107 Chemnitz, Germany}

\author{Benny B\"ohm}
\affiliation{Institute of Physics, Chemnitz University of Technology, Reichenhainer Strasse 70, 09107 Chemnitz, Germany}

\author{Sebastian Schneider}
 \affiliation{Dresden Center for Nanoanalysis, cfaed, Technische Universität Dresden, 01069 Dresden, Germany}
 
\author{Darius Pohl}
 \affiliation{Dresden Center for Nanoanalysis, cfaed, Technische Universität Dresden, 01069 Dresden, Germany}
 
\author{Bernd Rellinghaus}
 \affiliation{Dresden Center for Nanoanalysis, cfaed, Technische Universität Dresden, 01069 Dresden, Germany}

\author{Aladin Ullrich}
 \affiliation{Institute of Physics, University of Augsburg, Universitätsstrasse 1, D-86159, Augsburg, Germany}

\author{Manfred Albrecht}
\affiliation{Institute of Physics, University of Augsburg, Universitätsstrasse 1, D-86159, Augsburg, Germany}

\author{Jürgen Lindner}
\affiliation{Institute of Ion Beam Physics and Materials Research, Helmholtz-Zentrum Dresden-Rossendorf, Bautzner Landstrasse 400, 01328 Dresden, Germany}

\author{Nikolai~S.~Kiselev}
 \email{n.kiselev@fz-juelich.de}
 \affiliation{Peter Gr\"unberg Institute and Institute for Advanced Simulation, Forschungszentrum J\"ulich and JARA, 52425 J\"ulich, Germany}

\author{Olav~Hellwig}
 %\email{o.hellwig@hzdr.de}
\affiliation{Institute of Ion Beam Physics and Materials Research, Helmholtz-Zentrum Dresden-Rossendorf, Bautzner Landstrasse 400, 01328 Dresden, Germany}
\affiliation{Institute of Physics, Chemnitz University of Technology, Reichenhainer Strasse 70, 09107 Chemnitz, Germany}

\date{\today}

\begin{abstract}
We report on the controlled switching of domain wall (DW) magnetization in aligned stripe domain structures, stabilized in  [Co (0.44 nm)/Pt (0.7 nm)]$_X$ ($X = 48$, 100, 150) multilayers with perpendicular magnetic anisotropy. The switching process, induced by an external magnetic field, is monitored by measuring the evolution of the in-plane magnetization. 
%FS: I don't know what "is independent on the entire rest of the magnetic structure" was referring to -- that for a given aligned stripe state, the domain magnetization doesn't change for small IP-fields? Else, it should depend on the magnetic structure I think, e.g. if it's a bubble state or stripe state
We show that the remanent in-plane magnetization originates from the polarization of the Bloch-type DWs. With micromagnetic simulations, we reveal that the reversal of the DW polarization is the result of the emergence and collapse of horizontal Bloch lines within the DWs at particular strengths of the external magnetic field, applied opposite to the DW polarization. Our findings are relevant for DW-based magnonics and bubble skyrmion applications in magnetic multilayers.  
\end{abstract}

\maketitle
\section{Introduction}
% Magnetic multilayer heterostructures with perpendicular magnetic anisotropy (PMA) are \NSK{\sout{a system} synthetic materials}  of high interest due to the many possibilities of their applications, for example, in magnetic recording media~\cite{1}, nonvolatile logics \cite{2} and magnon-based data transfer devices \cite{3,3a,4}. 

Magnetic multilayers  (MLs) with perpendicular magnetic anisotropy (PMA) are synthetic materials with highly tunable properties. The accurate control of the magnetic properties in these heterostructures is realized by adjusting their design parameters, e.g. ML periodicity, ferromagnetic sublayer thickness and number of repeats. Over the years, these materials found wide use in many fields, such as data storage technology~\cite{1,Lambert}, nonvolatile logic~\cite{2}, spintronics~\cite{Mangin,Hirohata}, magnonics \cite{3,3a,4,4a} and biotechnology~\cite{Vemulkar_15}.

Particular interest concerns magnetic MLs with high quality factor, $Q = K_\mathrm{u}/K_\mathrm{d}$, which represents the ratio between the PMA energy density at in-plane saturation, $K_\mathrm{u}$, and magnetostatic energy density at out-of-plane saturation, $K_\mathrm{d}=\frac12\mu_0 M_\mathrm{s}^2$, where $\mu_0$ is the vacuum magnetic permeability and $M_\mathrm{s}$ is the saturation magnetization.
In MLs with $Q \ll 1$ the ground state is a single domain state with magnetization lying in the plane of the film.
In the case of strong PMA ($Q \gtrsim 1$) the competition between long-range magnetostatic interactions and short-range exchange interaction results in the formation of magnetic domains with ``up'' and ``down'' magnetization, which can form labyrinth patterns, aligned stripe domains or cylindrical magnetic bubble domains~\cite{8,9}.
The control of such magnetic states by external stimuli, such as magnetic field, electric current or temperature offers vast perspectives for fundamental research and applications. 

MLs with PMA have been reconsidered recently as a host materials for bubble-skyrmions, i.\,e. bubble domains with defined domain wall (DW) chirality in the presence of interfacial Dzyaloshinskii-Moriya interaction (iDMI). The iDMI is induced by asymmetric interfaces, such as Ir/Fe/Co/Pt or Pt/Co/Ir~\cite{2,34}.

The well-established MLs with symmetric interfaces, such as Co/Pt or Co/Pd, still offer rich perspectives, for example, for magnon-based data transfer applications~\cite{3,3a,4,4a,13}.
Particularly, aligned stripe domain structures stabilized in these MLs have been suggested as magnetically reconfigurable system for spin wave propagation~\cite{3,3a,4,4a,13}.
It has been demonstrated that the direction of the domains can be reconfigured using an external magnetic field or electrical current~\cite{13,14,15}. 
Besides the stripe domain direction, the magnetization orientation inside the DWs can be another degree of freedom to control the characteristics of the spin waves localized in the DWs~\cite{3a,4,4a}.

In this work, we report on internal switching of the DW polarity, induced by in-plane magnetic fields, in Co/Pt MLs with strong PMA.
Such switching is accompanied by the emergence of horizontal Bloch lines (HBL) inside the DWs of aligned stripe domains.
The concept of a HBL has been well-established for garnet films and other bubble materials~\cite{Malozemoff_79, 16}, however, to the best of our knowledge, has not been reported for magnetic MLs yet.  

Using vibrating sample magnetometry (VSM) we study the magnetization reversal inside the DWs by performing in-plane minor hysteresis loop series with the field applied parallel to the stripe domain long axis.
These measurements reveal a cascade of transitions leading to a step-like evolution of the in-plane remanent magnetization as a function of the previously applied field.
We argue that the observed behavior in our system generally applies for PMA MLs with symmetric interfaces. 
This argument is supported by the reproducibility of such a cascade of transitions in a sample series with different thicknesses.
Our study further comprises experimental measurements with magnetic force microscopy (MFM), Lorentz transmission electron microscopy (LTEM), as well as micromagnetic simulations, which allow us to reveal fine details of the magnetization reversal processes inside individual domain walls.

%These results will be interesting and inspiring for a wide community of researchers working on PMA MLs with symmetric and asymmetric interfaces.

The paper is organized as follows. We first outline the sample preparation and experimental techniques, including the method for the stripe domain alignment. Then we discuss the in-plane magnetization minor loops and the remanent magnetization behavior as a function of the previously applied field, along with supporting MFM and LTEM images. Finally, we show the results of micromagnetic simulations and provide a comprehensive picture of the magnetization reversal processes and the role of HBLs.
These results contribute to a fundamental understanding of DW magnetic polarization and switching in PMA MLs, inspiring their research and applications with symmetric and asymmetric interfaces.

\section{Experimental details}

\begin{figure}[t]
    \includegraphics[width=0.9\linewidth]{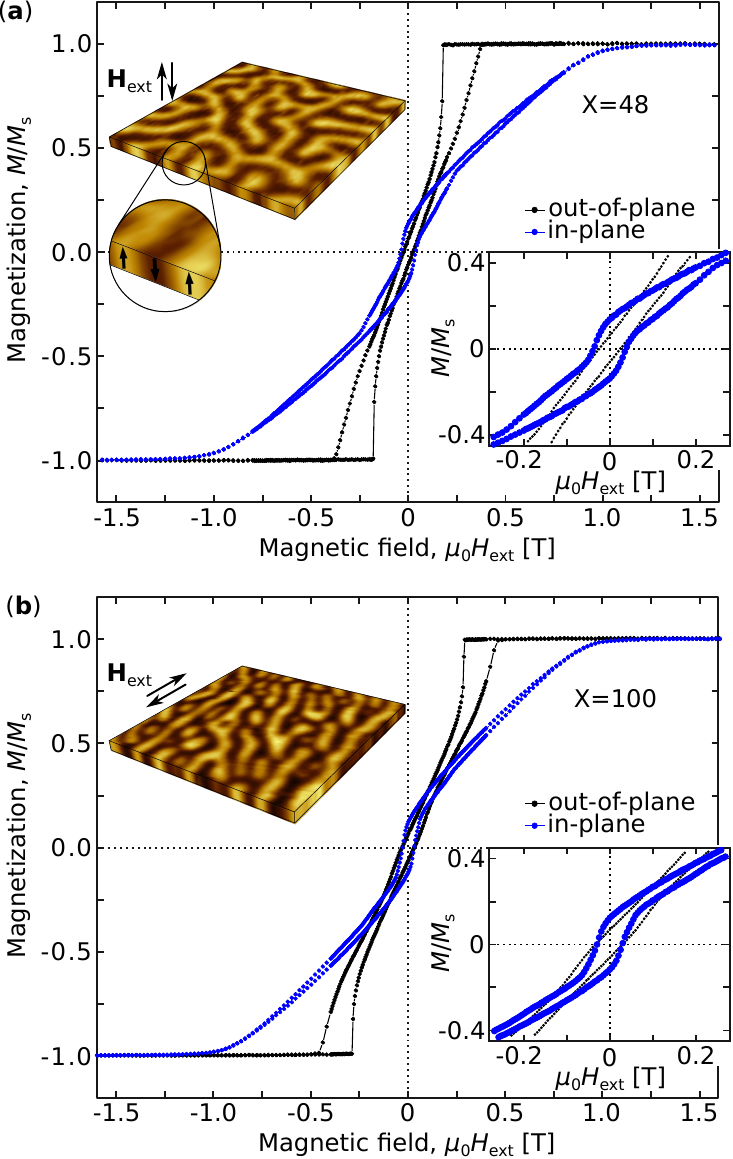}
    \caption{
Magnetic hysteresis loops measured with magnetic field applied perpendicular (black circles) and parallel (blue circles) to the [Co (0.44\,nm)/Pt (0.7\,nm)]$_X$ multilayer surface for $X = 48$ (a) and $X = 100$ (b). The bottom insets show an enlarged view of the corresponding hysteresis loop at small magnetic fields.
The top insets are schematic illustrations adapted from approximately $2\mu\mathrm{m}\times 2\mu\mathrm{m}$ sized MFM images for the multi-domain state at remanence, $H_\mathrm{ext}=0$, after out-of-plane (a) and in-plane (b) magnetic saturation. 
}
\label{fig:Fig1}
\end{figure}

The [(Co(0.44\,nm)/Pt(0.7\,nm)]$_X$, $X = 48$, 100, 150 ML films were fabricated at room temperature by DC magnetron sputter deposition at 0.4\,Pa Ar atmosphere in an ultrahigh vacuum system ATC 2200 from AJA International Inc. 
We used Si substrates with 100 nm thick thermally oxidized (SiO$_2$) layer. Prior to the multilayer deposition, a 1.5 nm Ta layer was grown for adhesion purposes. 
A subsequent 5\,nm or 20\,nm thick Pt buffer serves as a seed layer for inducing a preferred (111)-texture of the Co/Pt MLs, which supports larger and more uniform PMA. 
The ML samples were finally capped by 2\,nm Pt to avoid surface oxidation. 
Magnetic measurements were performed at ambient temperature on a commercial Microsense EZ7 vibrating sample vector magnetometer (VSM), equipped with an electromagnet, and a goniometer for angular dependent measurements. For magnetic domain imaging we used a Bruker Dimension Icon magnetic force microscope. 
All MFM images were recorded at room temperature and zero magnetic field. 

The domain wall magnetization imaging was performed using LTEM. 
For LTEM experiments the [(Co(0.44\,nm)/Pt(0.7\,nm)]$_{48}$ ML was fabricated on a silicon nitride TEM grid using a 5 nm Pt buffer layer in order to reduce absorption. The LTEM investigation was performed with a JEOL NEOARM TEM (JEM-ARM200F), operated at 200 keV beam energy, in the Lorentz mode. Images were taken with a Gatan OneView camera. No further treatment (like plasma cleaning) was applied to the samples.

The aligned stripe domain states in all the samples were prepared using a specific AC-demagnetization (ACD) routine, namely by alternating the magnetic field direction parallel to the sample surface with subsequent reduction of the field amplitude from 1600\,mT down to 2\,mT in $2\,\%$ steps. This routine brings the sample into a fully demagnetized state, which is characterized by zero remanence. 
The saturation magnetization of all samples was measured using the VSM and calculated to be $M_\mathrm{s}$ = 0.77 ± 0.07\,MA/m. The PMA constant was determined from the in-plane saturation field of a ML with identical structure as the samples investigated here, but with smaller X = 12 (yielding a square out-of-plane hysteresis). 
We obtained $K_\mathrm{u} \approx 0.6$\,MJ/m$^3$, leading to the value $Q \approx 1.6$, therefore the strong stripe domain criteria are met for all samples investigated here~\cite{Malozemoff_79, 16, 11, Virot}.

\section{Results}

\begin{figure*}[ht]
    \includegraphics[width=0.9\linewidth]{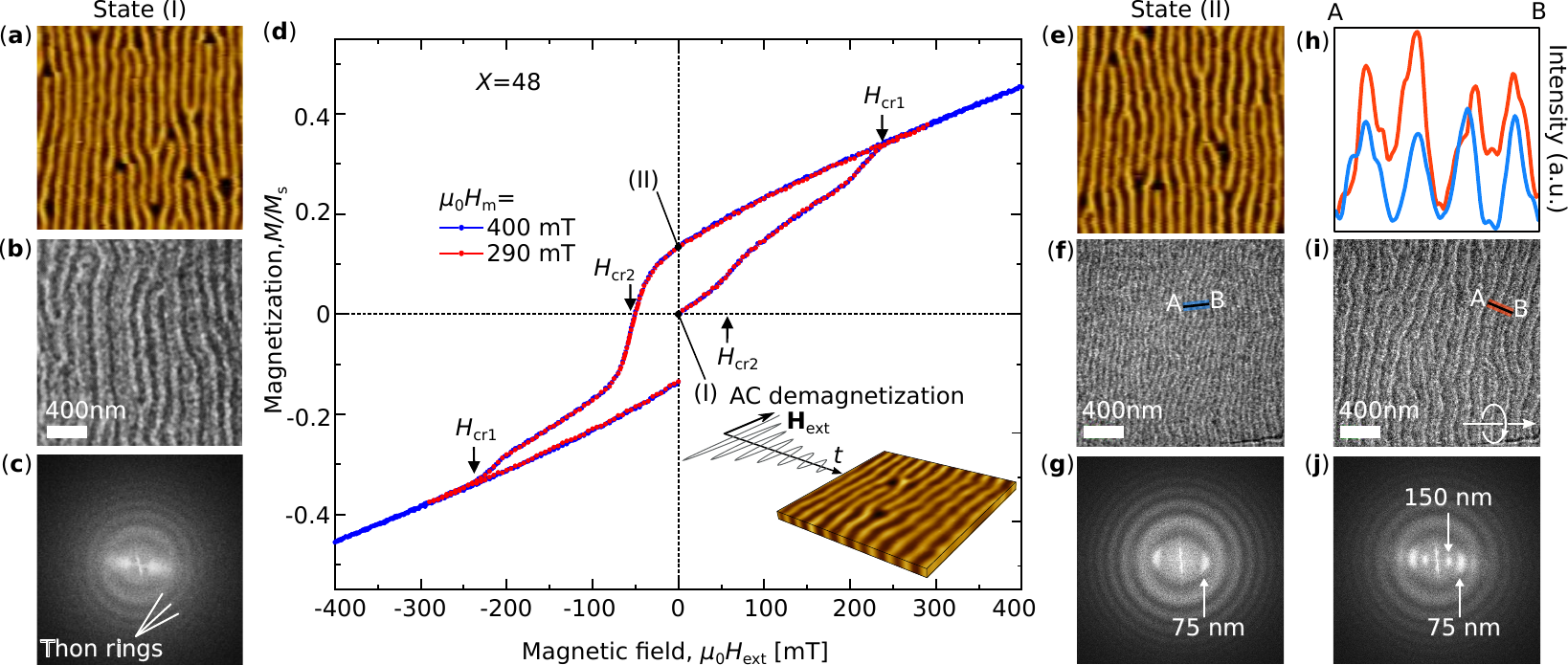}
    \caption{
(a) MFM image of the aligned stripe domain state and (b) LTEM magnetic contrast with corresponding (c) FFT of the magnetization inside the DWs in the [Co (0.44\,nm)/Pt (0.7\,nm)]$_{48}$ sample. Initially the sample is in an in-plane ACD state. (d) Minor hysteresis loops up to 290\,mT (red line) and 400\,mT (blue line). The field was applied parallel to the stripe domains, starting from the demagnetized state (I). 
The states (I) and (II) marked in the loops correspond to the domain states presented in (a), (b), and (e), (f), respectively.
(e) MFM image of stripe domains recorded at zero field after exposing the demagnetized sample to a field of 300\,mT, (f) corresponding LTEM magnetic contrast of the DWs and (g) its FFT.  (i) LTEM contrast and (j) corresponding FFT for an inclined sample region (due to the wavy nature of the membrane with MLs) with respect to the normal incidence of the electrons. 
In that case the LTEM contrast from the out-of-plane domains gets superimposed onto the LTEM contrast from the DWs. (h) intensity amplitudes of the LTEM contrasts in (f) - blue line and in (i) - red line. All MFM and LTEM images were recorded at zero magnetic field and at room temperature, but at different locations on the sample. 
\label{fig:Fig2}
}
\end{figure*}

% We recall studies from the late eighties, when the magnetic domain structures in garnet films were intensively investigated for magnetic memory applications~\cite{17,18,19,20}.
Before discussing our results in detail we briefly recall studies from the late eighties, when the magnetic domain structures in garnet films were intensively investigated for magnetic bubble memory applications~\cite{17,18,19,20}.
Back then it was demonstrated for the first time that in high-Q films the magnetization inside the DWs between aligned stripe domains can be switched by applying an in-plane magnetic field.
Moreover it was shown that the switching of the DW polarity does not affect the magnetic structure and stability of the stripe domains~\cite{17,18,19,20}.
Here, we show that a similar switching process takes place in magnetic MLs. However, the mechanism of the switching is different to the one proposed in Refs.[~\cite{17,18}].

Magnetic hysteresis loops for the [Co (0.44\,nm)/Pt (0.7\,nm)]$_X$ MLs are presented in Fig. 1(a) and (b) for $X = 48$ and $X = 100$, respectively. Both samples exhibit a similar magnetic response typical for MLs with PMA~\cite{5,6,7}. 
%
%At small magnetic fields, magnetization measured in the out-of-plane geometry shows almost linear dependence on the applied field, indicating gradual increase of reversed domain width with lowering \NSK{[maybe increasing?]} the field \NSK{\textbf{[The mission of this sentence is not entirely clear for me]}}. 
%
Magnetization measured in the out-of-plane geometry shows easy-axis behaviour, as expected for MLs with $Q \gtrsim 1$.
At remanence, the system evolves into a labyrinth-type domain structure with almost %FS: since there is a very small non-zero remanence
equal width of “up” and “down” domains, shown in the top-left inset of~Fig.~1(a). 
In the in-plane geometry, the magnetization shows hard axis behavior, however, with non-zero remanence (see the bottom right insets in Fig.~1(a) and (b)). 
There, magnetic domains evolve to a mixed state consisting of short and elongated stripe domains in combination with some bubble domains (top left inset of Fig. 1(b)), when decreasing the magnetic field from saturation to zero. 
The normalized remanent magnetization $m_\mathrm{r}$ = $M_\mathrm{r}$/$M_\mathrm{s}$ ($M_\mathrm{r}$ being the magnetization at zero field) decreases slightly with $X$ (see Table I), indicating a weak dependence on the sample thickness.

In the following, we demonstrate that the remanent magnetization measured parallel to the stripe domain long axis originates from the magnetization inside of Bloch-type DWs.
As shown in Fig.~\ref{fig:Fig2}(a), after the ACD protocol, the samples exhibit parallel stripe domains aligned along the magnetic field direction. 
%as this represents the energetically most favorable state at applied in-plane fields.
%
% The parallel alignment of the stripe domains to the external field direction is the result of the interaction of the magnetic field with the Bloch component of the stripe DWs.. 
%
In contrast to the case where we reduce the in-plane field directly from saturation to zero (Fig.~\ref{fig:Fig1}), now we observe a vanishing remanent magnetization (see the state marked by ``(I)'' in Fig.~\ref{fig:Fig2}(d)).

The aligned stripe domain structure in Fig.~\ref{fig:Fig2}(a) exhibits a domain period of 160 ± 10 nm. 
Figure~\ref{fig:Fig2}(b) shows the LTEM magnetic DW contrast for these stripe domains and Fig.~\ref{fig:Fig2}(c) is the corresponding fast Fourier transformation (FFT). 
%
% Irregular structure of the magnetic induction from the DWs is seen and expected, since the system is in a fully demagnetized state, out-of-plane (domains) as well as in-plane (domain walls). 
%
% The Bloch components of the DWs are randomly distributed along the two possible alignment directions, which is confirmed by the random periods shown in Fig. 2(b) and the broad peaks in the FFT in Fig. 2(c) from periods of 60-500 nm.
The irregular contrast of the LTEM image and the broad peaks in the FFT indicate a random alignment of the in-plane Bloch wall component of different DWs (along the two possible directions), which is expected after ACD.

\begin{figure*}[ht]
    \includegraphics[width=0.9\linewidth]{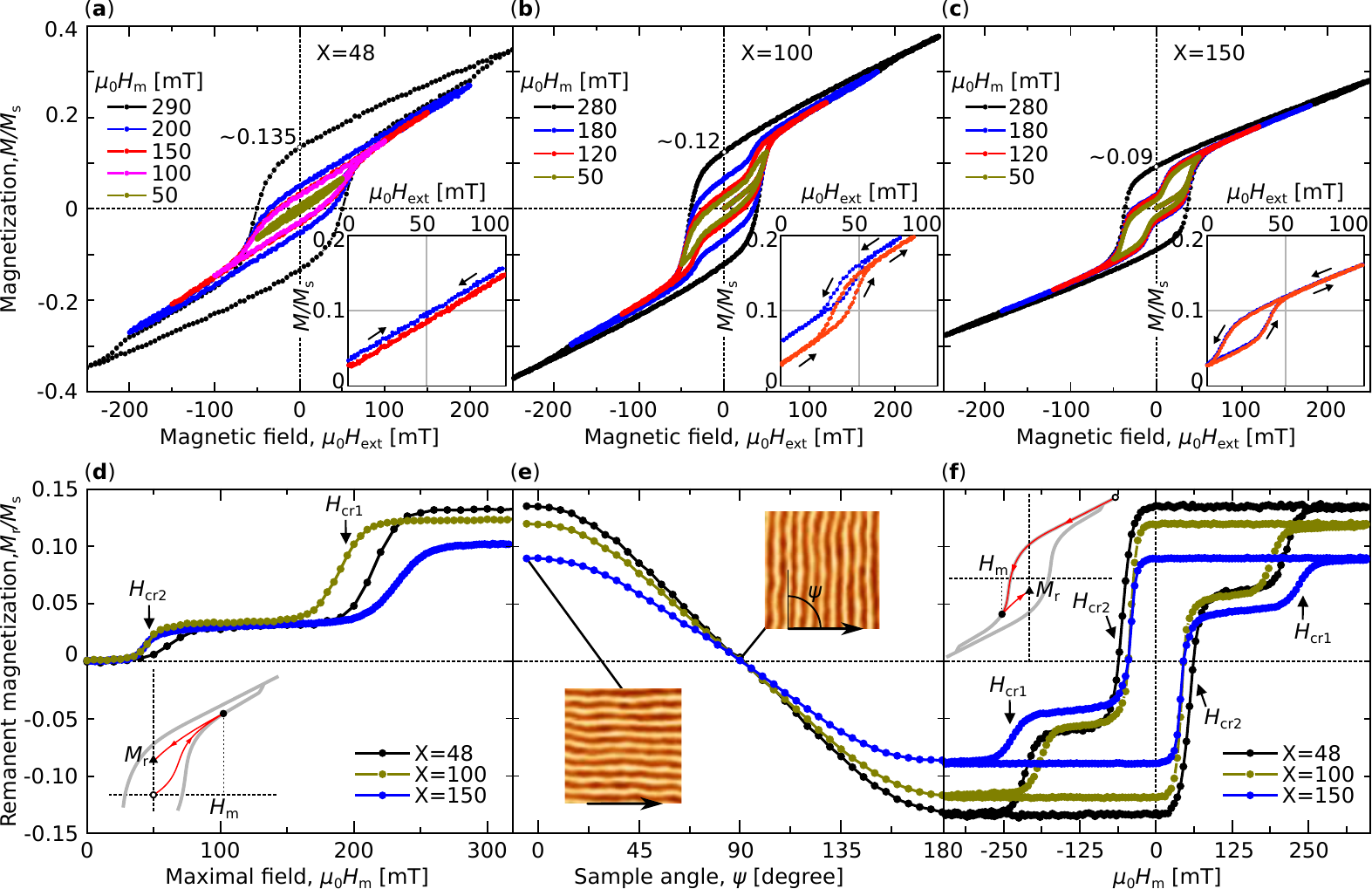}
    \caption{
Magnetic minor loop series with different maximum magnetic fields, $H_\mathrm{m}$, for [Co (0.44 nm)/Pt (0.7 nm)]$_X$ multilayers with (a) $X = 48$, (b) $X = 100$ and (c) $X = 150$. The field is applied parallel to the long axis of the stripe domains, which were created by previous in-plane ACD.
The insets in (a)-(c) show an enlarged view  of the minor loops near the inflection point associated with the critical field $H_\mathrm{cr2}$. 
(d) Normalized remanent magnetization as a function of the previously applied field, $\mu_0H_\mathrm{m}$, starting from the aligned stripe state stabilized by ACD. The inset shows in detail the initial state after ACD (hollow circle), the state at maximum field (solid circle) and final state at remanence (triangle), which is displayed in the main plot of (d).
(e) Angular dependence of the in-plane remanent magnetization of the aligned stripe domains, where $\psi=0$ corresponds to the direction parallel to the stripes. 
(f) Normalized remanent magnetization as a function of the maximum applied magnetic field, $\mu_0H_\mathrm{m}$, varied between $+300$\,mT and $-300$\,mT and back.
In contrast to (d), in these measurements, the initial state is an in-field state, see the hollow circle in the inset. 
Note that all data points in (d), (e) and (f) are taken at zero magnetic field, and all three graphs have identical $y$-axis scale. 
\label{fig:Fig3}
}
\end{figure*}

In a second step, we then apply a moderate in-plane field along the stripe domain direction in order to align all the Bloch components of the DWs into one of the two possible orientations and then reduce the field back to zero. 
We now pick up a significant remanent in-plane moment ($m_\mathrm{r}$ = 0.135) along the stripe domain direction as visible for state “(II)” in Fig. 2(d). 
The MFM image, shown in Fig. 2(e), has not changed in any significant way and the domains are still in a fully aligned state with the period of 160 ± 10 nm. 
However, the LTEM image (Fig. 2(f)) shows now, at normal incidence of the electrons, where we are only sensitive to the projected in-plane magnetic induction, hence in-plane Bloch wall components, a much narrower stripe pattern with a well-defined period of {75 ± 5\,nm} (Fig. 2(g)). 
This confirms that now all the Bloch components of the DWs are pointing in the same direction and thus result in a period of {75 ± 5\,nm}, which is coinciding with half of the domain pattern period obtained from the MFM images (within the error bars determined by the FFT peak broadening).

%\FS{[Comment: Instead of calibration differences [both tools should be sub-nm exact] I would suggest small tilting of the membrane and non-perfect parallel alignment of the stripes (with stacking faults) as an explanation (the FFT is not a point but has some broadness), together with the limited field of view increasing the uncertainty]}
%
%\RS{[Comment: Very good and important point! Actually, when I decrease 'statistics' and account only perfectly aligned region, I always end up with 75 nm. 80 nm and larger statistical error comes when I consider 'full MFM image'. Fabian, I'm fully agree, could you, please 'excuse the MFM'? The LTEM pictures shows very small area of the membrane. It can not be tilted, since this region is picked as a region at normal incidence of electrons. Also, according to the LTEM FFT peak linewidth, we should put error bar for the LTEM periodicity.]}
%
%Finally we tilt the sample in order to have the electrons penetrate the sample with a small angle to the surface normal and thus create also some additional Lorentz contrast from the domain themselves (Fig. 2(j)). 
%

Figure~\ref{fig:Fig2}(i) shows the magnetic contrast from an inclined region of the membrane with respect to the electron beam. In that case electrons penetrate the sample with an angle to the surface normal and thus experience some additional deflection due to the in-plane magnetic induction components of the domains themselves.
In such a geometry, we see now both periods, the 150\,nm period from the domains, superimposed with the 75\,nm period from the “artificially aligned” DWs (Fig.~\ref{fig:Fig2}(j)).
This is also clearly visible in the comparison of the LTEM image line profiles of normal incidence (Fig.~\ref{fig:Fig2}(f), blue) versus inclined incidence (Fig.~\ref{fig:Fig2}(i), red) regions, displayed in Fig.~\ref{fig:Fig2}(h). Indeed, the red curve exhibits an additional sinusoidal 150\,nm-modulation as compared to the blue curve.
The original AC demagnetized sample does not show any visible periodicity of 75\,nm. As a result we can conclude that the non-zero remanent in-plane moment measured along the stripe direction originates from the polarized domain wall magnetization.

Having confirmed that the remanent in-plane magnetization originates from the Bloch-type DW polarization, we studied characteristics of the DW magnetic switching. In the following, if not otherwise mentioned, we present and discuss data for in-plane magnetic fields applied parallel to the stripe domain long axis.
The minor loop series for MLs with $X = 48$, 100 and 150, are displayed in Fig. 3(a), (b) and (c), respectively. 
With increasing the magnetic field, the magnetization increases and reveals a hysteretic behavior with an opening between ascending and descending branches of the minor hysteresis loops. 
As soon as the maximum magnetic field, $H_\mathrm{m}$, in the minor loops series overcomes $H_\mathrm{cr1}$ ($\mu_0H_\mathrm{cr1}$ = 240 mT for X = 48, see Fig. 2(d) and Fig. 3(a)) the descending branch of the loop shows constant remanence, which is independent of $H_\mathrm{m}$. 
The positive slope above $H_\mathrm{cr1}$ results from the continuous and reversible rotation of the magnetization inside the domains towards the external magnetic field direction~\cite{17,18,6,22,23}. 
All samples show switching from a zero point demagnetized state to a finite remanent magnetization state at $H_\mathrm{m} > H_\mathrm{cr1}$.
The critical field $H_\mathrm{cr1}$ and the remanent magnetization for all MLs are summarized in Table I. 
%
%\FS{[Comment: Order changed (I find it confusing/not necessary to jump from 3c to 3e back to 3d) and the discussion of 3e shortened]}

For small fields $H_\mathrm{m} \leq H_\mathrm{cr2}$ ($\mu_0H_\mathrm{cr2}$ = 50\,mT for $X = 48$), on the other hand, the magnetization shows a linear and reversible response to the field.
Accordingly, the remanent moment remains zero, as is also visible in Fig.~\ref{fig:Fig3}(d), which presents the plot of $m_\mathrm{r}$ as a function of $H_\mathrm{m}$.
The remanent magnetization for all samples shows a characteristic double step behavior as a function of $H_\mathrm{m}$ until $H_\mathrm{m}$ overcomes the critical field $H_\mathrm{cr1}$ (see the plot in Fig.~\ref{fig:Fig3}(d)). 
Another plateau appears at $H_\mathrm{cr2}$ < $H_\mathrm{m}$ < $H_\mathrm{cr1}$ (Fig. 3(d)), corresponding to approximately 20$\,\%$ - 30$\,\%$ of the maximum remanent moment. 
This plateau represents an intermediate stable DW polarization remanent state. 
%This plateau means that the DW polarization in the remanent ground state does not change significantly regardless of the magnetic field strength previously applied (within above specified bounds). 
%
The associated energy barriers for the DW magnetization reversal at the vicinity of $H_\mathrm{cr1}$ and $H_\mathrm{cr2}$ are modeled and discussed below.
Note that the irreversibility point in the major in-plane magnetic hysteresis loops in Fig.~\ref{fig:Fig1}(a) and (b) occurs at a higher field of approximately 720\,mT, showing that domain deformation requires larger magnetic field energies as compared to DW switching.

We further investigate the direction of the total net in-plane remanent magnetization by measuring $m_\mathrm{r}$ for different in-plane angles $\psi$ between the stripe domain long axis and the VSM pick-up coil axis, see Fig.~\ref{fig:Fig3}(e). Prior to this measurement, the sample was brought into a stripe domain state by ACD, then a field of 300\,mT (which is above $H_\mathrm{cr1}$ for all samples) was applied along the stripe axis. For $\psi=0^\circ$, the moment is measured parallel to the previously applied field, which gives the maximum signal from the polarized DWs. For $\psi=90^\circ$, on the other hand, the signal vanishes.
This shows that the net in-plane moment in our system is indeed parallel to the stripe DWs, and can be identified as the DW Bloch components polarized in one direction by the previously applied magnetic field protocol.

%This shows that indeed only the net in-plane moment in our system is parallel to the stripe DWs, namely the DW Bloch components polarized in one direction by the previously applied magnetic field.

So far we discuss the DW magnetization behavior starting from the fully demagnetized state after the ACD protocol to the polarized state. Figure~\ref{fig:Fig3}(f) plots the remanent magnetization for $\mu_0H_\mathrm{m}$ varied between -300 mT and +300 mT, thus the plot represents full magnetization reversal inside the DWs.
The double step behavior is also evident, with critical fields ($H_\mathrm{cr1}$ and $H_\mathrm{cr2}$) identical to the ones defined in Fig. 3(d) and as given in Table I. 
The DW magnetization changes sign and shows a plateau at $H_\mathrm{cr2} < H_\mathrm{m}< H_\mathrm{cr1}$, however, this time the magnetic polarization reaches about 40$\,\%$ - 50$\,\%$ of its maximum.
 
 The minor hysteresis loops for our ML series show a systematic trend with the sample thicknesses (X): the loop opening in between the two critical field values $H_\mathrm{cr2}$ and $H_\mathrm{cr1}$ is narrowing down with increasing ML thickness X as seen when comparing minor loop series for $X = 48$, 100 and 150 in Fig.~\ref{fig:Fig3}(a), (b) and (c). 
 Furthermore, the DW magnetization of X = 100 and 150 samples shows partial exchange spring like behavior until the magnetic field exceeds $H_\mathrm{cr1}$. The exchange spring behavior was reported for 2D-domain walls in hard/soft bilayer exchange spring magnets~\cite{26,28,29}, where the point of irreversibility is only reached when the DW penetrates into the hard layer.  
 In thick (above 1 $\mu$m) garnet films this behavior was clearly demonstrated and referred to as a first jump in the double-jump magnetization reversal of the stripe DWs~\cite{18}.

Since the in-plane remanent magnetization is attributed to the polarization of the DW Bloch component, it becomes possible to estimate an effective DW width (corresponding to a linear Bloch DW profile without N\'{e}el caps), $\Delta$, for the aligned stripe domain configuration. A simple equation can be used~\cite{Virot,6}:
\begin{equation}
\Delta =2\lambda\cdot{m_\mathrm{r}}.
\label{M_R0}
\end{equation}
The stripe domain width, $\lambda$, is determined from the MFM images. The effective domain wall width, presented in Table I, is independent of the ML thickness. This is expected from the models for stripe domains in high-$Q$ materials with the thicknesses above the characteristic critical thickness~\cite{Virot,12}. Since the DW width does not change with the sample thickness, magnetic MLs with larger domain periodicity should exhibit smaller remanent magnetization. This explains why $m_\mathrm{r}$ decreases with $X$ in our sample series.

\begin{table}[ht]
\caption{\label{Tab} The total magnetic film thickness $t$, normalized remanent magnetization, $m_\mathrm{r}=M_\mathrm{r}/M_\mathrm{s}$, DW saturation field, $\mu_0 H_\mathrm{cr1}$, DW magnetic activation field, $\mu_0 H_\mathrm{cr2}$, averaged stripe domain width, $\lambda$, and effective DW width, $\Delta$, for [Co (0.44\,nm)/Pt (0.7\,nm)]$_X$ multilayers with $X =$ 48, 100 and 150.}
\begin{tabular}{p{0.09\textwidth}p{0.115\textwidth}p{0.115\textwidth}p{0.115\textwidth}}
\hline
\hline
                       & X=48   & X=100  & X=150    \\
\hline
$t$                    & 55\,nm         & 114\,nm        & 171\,nm  \\
\hline
$m_\mathrm{r}$          & 0.135         & 0.12          & 0.09  \\
\hline
$\mu_0 H_\mathrm{cr1}$  & 240\,mT        & 200\,mT        & 250\,mT  \\
\hline
$\mu_0 H_\mathrm{cr2}$  & 50\,mT         & 40\,mT         & 37\,mT   \\
\hline
$\lambda$               & 80$\pm$5\,nm   &91$\pm$5\,nm    &115$\pm$5\,nm  \\
\hline
$\Delta$                & 22$\pm$2\,nm   &22$\pm$2\,nm    &20$\pm$2\,nm  \\
\hline
\hline
\end{tabular}

\end{table}

\section{Micromagnetic simulations}

\begin{figure*}[ht]
    \includegraphics[width=0.9\linewidth]{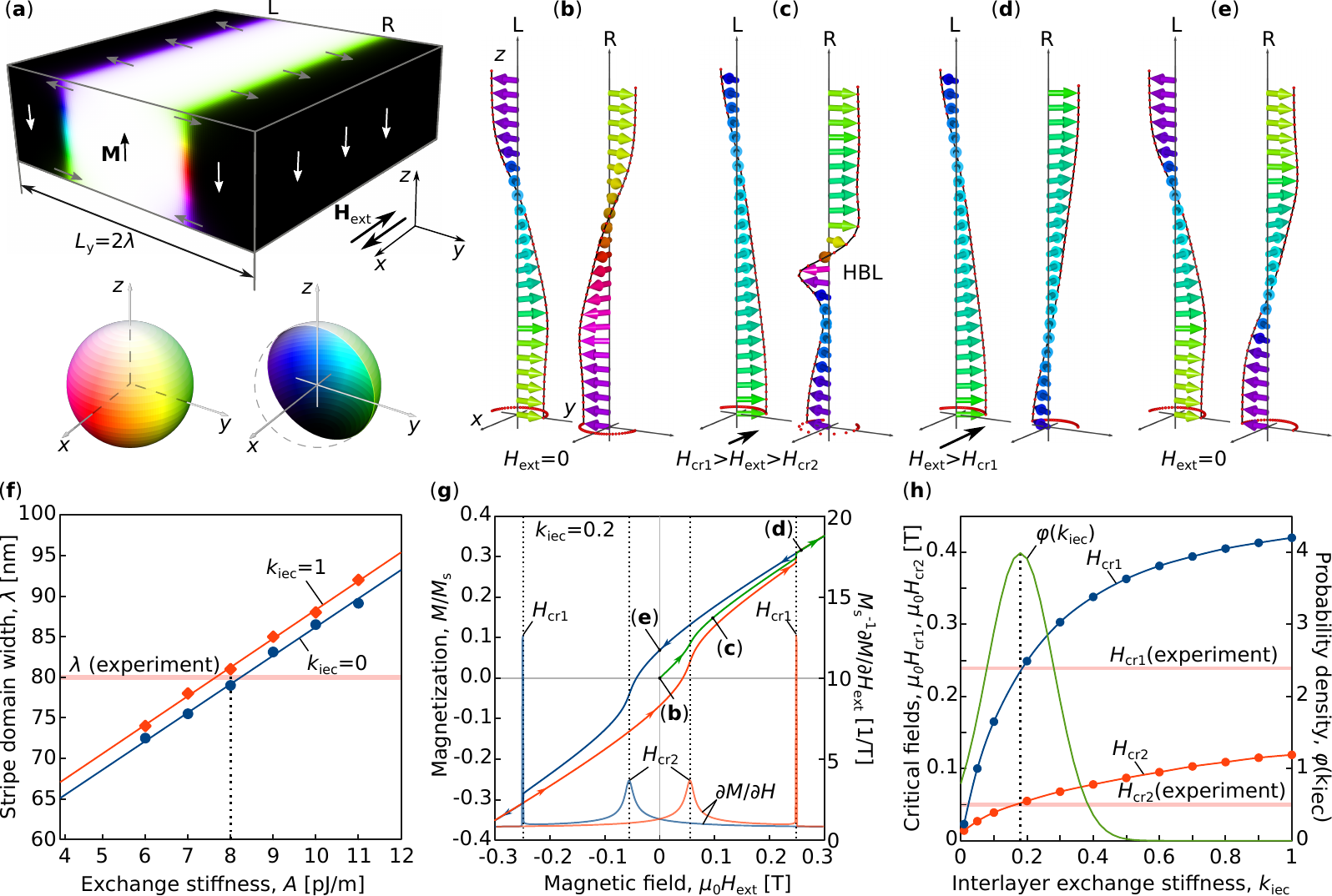}
    \caption{
    (a) The simulated domain and equilibrium magnetization distribution at zero external field for the $X = 48$ ML. The unit vector field for magnetization is represented with the color code displayed by the inset below. 
    (b)-(e) The magnetization along the thickness of the sample inside the left (L) and the right (R) domain walls, see (a), at different external magnetic field applied antiparallel to the $x$-axis. 
    (f) The equilibrium stripe domain width calculated for different values of exchange stiffness constant $A$ and coefficient of interlayer exchange coupling, $k_\mathrm{iec}$, see Eq.~\eqref{w_ex}.
    (g) The representative magnetization curve calculated on the domain shown in (a) by energy minimization at each field.
    The green line is the virgin magnetization curve starting from the demagnetized state depicted in (a).
    The blue and the red lines are the descending and ascending branches of the minor magnetization loop, respectively.
    The peaks of the differential susceptibility $\partial M/\partial H$ shown for descending (blue) and ascending (red) branches indicate the two transitions.
    (h) The dependence of the critical fields $H_\mathrm{cr1}$  and $H_\mathrm{cr2}$ on the interlayer exchange coupling.
    The vertical dashed line indicates the optimal value, $k_\mathrm{iec}=0.18$, at which the theoretical fields match the experimental ones.
    The green curve shows a normal distribution for $k_\mathrm{iec}$ with mean $k_\mathrm{iec}=0.18$ and assuming a variance $\sigma^2=0.01$.
    }
\label{fig:T1}
\end{figure*}

To explain the physical nature of the magnetization curves and shed some light on the process taking place inside the domain wall during the magnetization reversal, we utilize micromagnetic simulations performed with mumax code~\cite{mumax}.
The main results of the numerical simulations are summarized in Fig.~\ref{fig:T1}.
The simulated domain has a shape of a cuboid with the thickness $L_\mathrm{z}=52.8$\,nm, which is approximately the magnetic thickness of the sample with $X=48$.
In the $xy$-plane the sample has a rectangular shape with the size fitting the equilibrium period of stripe domains, $L_\mathrm{x}=L_\mathrm{y}=2 \lambda$ and contains two DWs.
We use periodical boundary conditions in the $xy$-plane and assume that the magnetic field is parallel to the $x$-axis, see Fig.~\ref{fig:T1}(a).
In the simulations we use the experimentally estimated values for saturation magnetization, $M_\mathrm{s}=0.775$\,MA/m, and out-of-plane anisotropy constant, $K_\mathrm{u}=0.6$\,MJ/m$^3$, for details see the listing of the mumax script provided in Appendix A. 
We assume that the exchange coupling between Co layers is weaker than the direct exchange within the layer, such that the overall Heisenberg exchange becomes anisotropic.
In the micromagnetic approximation, the corresponding energy density term can be written as
\begin{equation}
w_\mathrm{ex}=A
%\int\limits_{V} 
\sum_i\!\left[
\left( \dfrac{\partial m_i}{\partial x}\right)^{\!2}\!+\!
\left( \dfrac{\partial m_i}{\partial y}\right)^{\!2}\!+
k_\mathrm{iec}\left( \dfrac{\partial m_i}{\partial z}\right)^{\!2}
\right],
%\mathrm{d}\boldsymbol{r}
\label{w_ex}
\end{equation}
where the summation runs over $i=x,y,z$. The constant $A$ is the exchange stiffness in the plane of the film and $\mathbf{m}=\mathbf{M}/M_\mathrm{s}$ is the magnetization unit vector field. 
The unitless constant $k_\mathrm{iec}$ in \eqref{w_ex} defines the strength of the exchange coupling between Co layers across the Pt layers.
For exchange decoupled layers, $k_\mathrm{iec}=0$, and in approximation of isotropic media, $k_\mathrm{iec}=1$.
We estimated the constants $A$ and $k_\mathrm{iec}$ from the fit of the experimental data for the domain size and critical fields $H_\mathrm{cr1}$ and $H_\mathrm{cr2}$.
Figure~\ref{fig:T1}(f) shows the dependence for the equilibrium domain size, $\lambda$, as a function of exchange stiffness, $A$, calculated for two limiting cases $k_\mathrm{iec}=0$ and 1.
We found the best agreement with the experimentally estimated $\lambda=80$\,nm for $X=48$ at $A=8$\,pJ/m, which represents a reasonable value for  Co/Pt MLs~\cite{Shim}.
The dependence of $\lambda$ on $k_\mathrm{iec}$ is too weak to allow for an estimation of $k_\mathrm{iec}$.
Instead, we performed a series of simulations for magnetization reversal curves with different $k_\mathrm{iec}$. 
The representative magnetization curve for $k_\mathrm{iec}=0.2$ is shown in Fig.~\ref{fig:T1}(g).
The profile of the magnetization inside the left (L) and right (R) DWs across the thickness are depicted in (b)-(e) for the different in-plane field values; the corresponding states are also marked in the magnetization curves (g).
In the initial demagnetized state the two DWs have opposite magnetization in the Bloch part of the DW (b).
The averaged in-plane component of magnetization projected onto the $x$-axis is zero in this case.
With increasing $H_\mathrm{ext}$ the magnetization inside the DW is rotated, tending to be aligned with the field direction.
For small fields $H_\mathrm{ext}<H_\mathrm{cr2}$ the profile of the magnetization changes insignificantly and is not shown here.
At the external field approaching $H_\mathrm{cr2}$ the magnetization profile changes abruptly (c).
In the R-DW, with the magnetization in the middle plane pointing against the external field, we observe the formation of a well-localized area with strongly twisted magnetization to which we refer as a horizontal Bloch line (HBL)\cite{Malozemoff_79}.
The emergence of the HBL leads to an abrupt increase of magnetization in the direction of the applied magnetic field, which results in the bending of the magnetization curve at $H_\mathrm{ext}=H_\mathrm{cr2}$.
Since the state with the HBL has a strong gradient of magnetization its energy is much higher than the energy of the DW with magnetization aligned with the field.
There is always a critical field $H_\mathrm{cr1}$ above which the HBL abruptly switches to the field-aligned state (d).

According to our observations, the switching occurs via nucleation of a pair of singularities -- Bloch points with opposite index.
The Bloch points run along the domain wall in opposite directions because the area between them represents the field-aligned state with lower energy.
The switching process finishes when the Bloch points reach the physical boundaries of the sample or meet other Bloch points with opposite index and annihilate. 
The dynamics of the Bloch points and the mechanism of their nucleation/annihilation are out of the scope of this work and will be discussed elsewhere in more detail.

When the external field reaches the critical value $H_\mathrm{cr1}$ and is then reduced to zero, the remanent magnetization exhibits a nonzero value because both DWs are now magnetized in the same direction (e).
With the increasing field in opposite direction, see the blue curve in (g) for negative $H_\mathrm{ext}$, the system undergoes the same sequence of transitions: first into the state with HBL at $|H_\mathrm{ext}|=H_\mathrm{cr2}$ and after that into the field-aligned state at $|H_\mathrm{ext}|=H_\mathrm{cr1}$.
Note that this time, the transitions take place in both domain walls simultaneously.
The two peaks in the differential susceptibility, $\partial M/\partial H_\mathrm{ext} $ , allow to identify the exact values of the critical fields.   
The dependence of $H_\mathrm{cr1}$ and $H_\mathrm{cr2}$ on the interlayer exchange coupling, $k_\mathrm{iec}$, are shown in Fig.~\ref{fig:T1}(h).
For $k_\mathrm{iec}=0.18$, the theoretically determined $H_\mathrm{cr1}$ and $H_\mathrm{cr2}$ are maximally close to the experimentally measured values. 
With that we conclude that with $A=8$\,pJ/m and $k_\mathrm{iec}=0.18$ our theoretical model describes the main features of the physical processes taking place in the experiment.

\begin{figure}%[ht]
\includegraphics[width=0.9\linewidth]{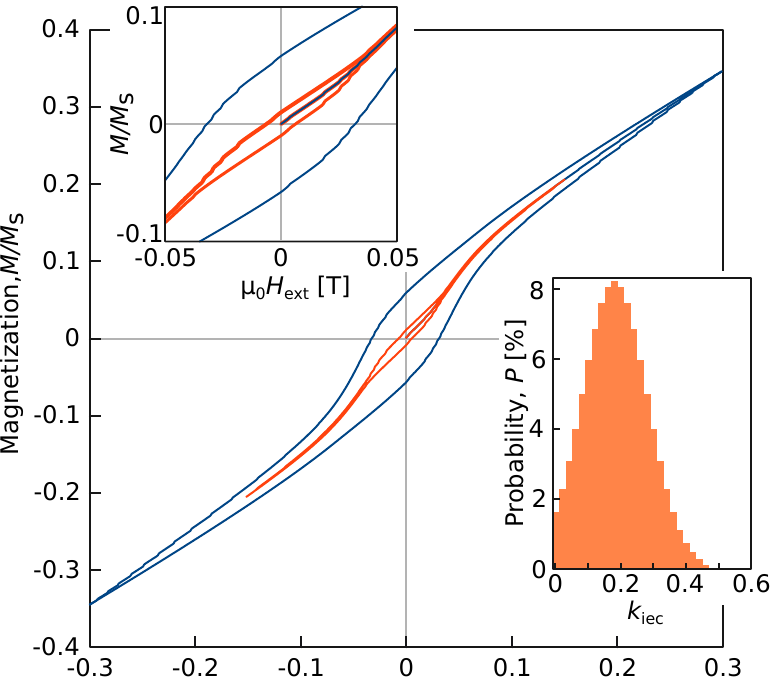}
\caption{
The minor loop magnetization curve above (blue) and below (red) the $H_\mathrm{cr1}$, calculated assuming a normal distribution of exchange coupling over the $X=48$ - sample. 
The top inset shows an enlarged view of the magnetization curves near remanence.
The bottom inset displays the probability distribution for the parameter $k_\mathrm{iec}$ representing the interlayer exchange coupling between adjacent Co layers.
The mean of the distribution corresponds to $k_\mathrm{iec}=0.18$.
}
\label{fig:T2}
\end{figure}

\begin{figure*}[ht]
\includegraphics[width=0.9\linewidth]{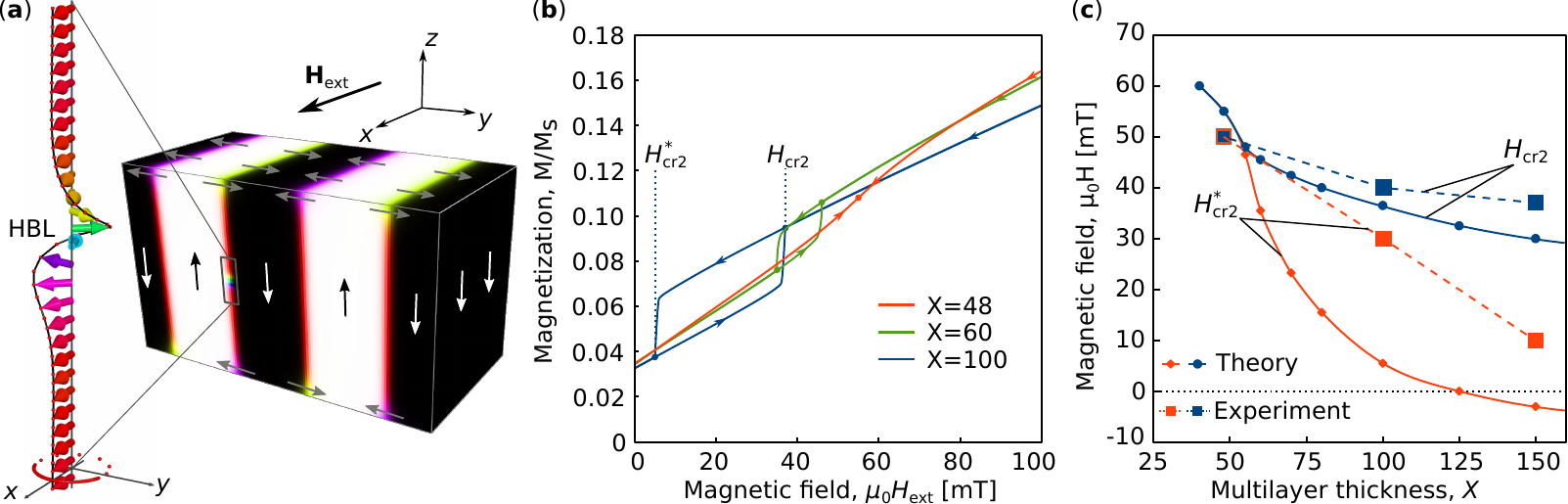}
\caption{
(a) The simulated domains for $X=100$ containing two periods of stripe domains separated by three DWs polarized along the external field and one DW with HBL. 
(b) The minor magnetization loops for MLs of different thickness simulated for the system (a). 
(c) The dependence of the critical fields $H_\mathrm{cr2}$ and $H^{*}_\mathrm{cr2}$ defining the opening of the minor loop as functions of the ML thickness.
}
\label{fig:T3}
\end{figure*}

The above theoretical model assumes a perfect crystal structure without any defects, grain boundaries, and other variations of the coupling constants inevitably present in samples prepared by magnetron sputter deposition. 
Because of that, the shapes of the theoretical and experimental magnetization curves differ in some details.
For instance the magnetization at $H_\mathrm{cr1}$ in the theoretical curve Fig.~\ref{fig:T1}(g) exhibits an abrupt jump while, in the experimental curves it shows a nearly continuous transition.
Moreover, the ideal crystal model can not explain the opening of minor magnetization loops seen in Fig.~3(a)-(c).  
Below we show that these discrepancies can be omitted at least on a qualitative level, assuming that the multilayer has a certain distribution of the interlayer exchange coupling, $k_\mathrm{iec}$, over the film area.
The green curve, $\varphi(k_\mathrm{iec})$, in Fig.~\ref{fig:T1}(h) shows the probability density function for a normal distribution with the  mean $\mu=0.18$ and variance $\sigma^2=0.01$.
The inset in Fig.~\ref{fig:T2} shows the probability $P$ as a function of $k_\mathrm{iec}$ on a discrete mesh with $\Delta k_\mathrm{iec}=0.02$.
The probability $P$ here means the percentage of stripes that belong to the structural region with a particular $k_\mathrm{iec}$.
For each value of $k_\mathrm{iec}$ with the step $\Delta k_\mathrm{iec}$ we calculated two minor loop magnetization curves with the maximal fields $\mu_0 H_\mathrm{m}= 0.3$\,T and 0.15\,T.
The averaged magnetization at each value of the external field has been calculated by
\begin{equation}
M(H_\mathrm{ext})=\sum\limits_{i}P_i\cdot M_i(H_\mathrm{ext}),
\label{M_av}
\end{equation}
where summation runs over all intervals of $k_\mathrm{iec}$ and $P_i$ stands for the probability for the $i$-th interval.
The averaged magnetization curves are shown in Fig.~\ref{fig:T2}.
Since for small $k_\mathrm{iec}$ the critical field $H_\mathrm{cr1}$ is small and the probability for the transition into the field-aligned state is nonzero, the resulting magnetization curves in both cases for $\mu_0 H_\mathrm{m}= 0.3$\,T and 0.15\,T have nonzero remanent magnetization. 
For $\mu_0 H_\mathrm{m}= 0.3$\,T the magnetization curve near $H_\mathrm{cr1}$ has a quite smooth behavior and fits the shape of the experimental curves better than the curve of Fig.~4(g).

Besides variations of interlayer exchange coupling in the real samples other parameters as for instance anisotropy strength and anisotropy direction might have their own distributions.
Because of that, the above simulations do not provide an exact quantitative agreement with the experimentally measured magnetization curves.
Nevertheless, these simulations clearly reveal that the presence of certain types of defects may trigger the transition from a DW with HBL into a field-aligned state even at a low magnetic field, $H_\mathrm{ext}\ll H_\mathrm{cr1}$.

\section{Discussions}

From the quantitative agreement of the critical fields and qualitative agreement of the experimental and theoretical magnetization curves, we conclude that the most realistic picture of the physical processes taking place during the application of in-plane magnetic fields along the stripe domains looks as follows.
The critical field $H_\mathrm{cr1}$ corresponds to the transition of all DWs into the field-aligned state. 
This explains the maximal remanent magnetization after the application of an external field exceeding $H_\mathrm{cr1}$. 
However, the transition of some individual DWs into the energetically favorable field-aligned state takes place even at lower external fields. 
The defects distributed over the volume of the sample, weak variations of exchange or anisotropy, as well as thermal fluctuations represent triggers for such transitions.
These events are rare at weak fields below $H_\mathrm{cr2}$ because the  energy barrier is larger. % 
This explains the first plateau in the $M_\mathrm{r}(H_\mathrm{m})$ curve, Fig.~\ref{fig:Fig3}(d).
When $H_\mathrm{m}$ approaches the critical field $H_\mathrm{cr2}$, the HBLs are formed in the DWs with opposite magnetization to the external field.
In the AC demagnetized state, approximately half of the DWs belong to this category, resulting in a total $M_\mathrm{r}=0$.
For an ideal, defect-free crystal, the DWs with HBL can survive up to $H_\mathrm{cr1}$, but due to the presence of defects, some portion of the DWs switches to the field-aligned state as soon as HBLs are formed (at about $H_\mathrm{cr2}$).
That happens because of the high gradients of magnetization in the DWs with HBL, which makes them sensitive to the presence of the defects. 
%
%\FS{\sout{Only a small amount of the domain walls with HBL and the close vicinity to the defects are switched at about $H_\mathrm{cr2}$, while their majority survive up to $H_\mathrm{cr1}$.}}

Denoting by $n_+$ and $n_-$ the number of DWs polarized along, respectively against, the previously applied field, and assuming that the total number of DWs $n_++n_-$ is constant, the remanent in-plane magnetization along the stripe axis reads.
\begin{equation}
\dfrac{M_\mathrm{r}}{M_\mathrm{s}}=\dfrac{n_{+}-n_{-}}{n_{+}+n_{-}}\cdot \dfrac{M^\mathrm{max}_\mathrm{r}}{M_\mathrm{s}},
\label{M_R}
\end{equation}
where $M^\mathrm{max}_\mathrm{r}$ is the maximal remanent magnetization.
From the relative height of the plateau in the $M_\mathrm{r}(H_\mathrm{m})$ curve (Fig.~\ref{fig:Fig3}(d)) and Eq.~\eqref{M_R} we estimated that for $X=48$ about $23\,\%$ of the antiparallel to the field polarized DWs undergo a switching at $H_\mathrm{cr2}$.
%
%\FS{$46\,\%$} of the domain walls\FS{\sout{, with respect to the total number of domain walls,}} undergo a switching at $H_\mathrm{cr2}$.
%This, in turn, corresponds to a switching of just \FS{$23\,\%$} of the antiparallely polarized DWs.
%
%\FS{\sout{The rest $94\%$ of the domain walls with HBL are switched only near $H_\mathrm{cr1}$}\small[Comment: The percentages were mixed up, or? The switches of all domain walls have to be half the percentage of the switches of specific aligned ones]}
%
The collective switching of all remaining antiparallel DWs into the field-aligned state at about $H_\mathrm{cr1}$ leads to the second large step in the curve shown in Fig.~\ref{fig:Fig3}(d).
With increasing the thickness, the relative number of the DWs undergoing this transition is almost unchanged.
For $X=100$ and $X=150$ we estimate that $\sim 27\,\%$  and $\sim 30\,\%$ of the DWs respectively, switch at $H_\mathrm{cr2}$.
The negligible thickness dependence is not surprising as the MLs were synthesized under almost identical conditions, and the small number of switching events at low field indicates a high quality of the samples.
Thereby, at the critical field $H_\mathrm{cr2}$ two processes take place: the appearance of HBL in the DWs with opposite to the field magnetization and partial switching of such DWs into the field-aligned state due to the presence of defects.

An interesting aspect revealed by the micromagnetic simulations is the hysteretic behavior of the DWs with HBL, as illustrated in Fig.~\ref{fig:T3}.
For a stripe domain state with its micromagnetically determined equilibrium period, we performed simulations starting with the state at a magnetic field, $\mu_0 H_\mathrm{ext}=180$\,mT, being in-between $H_\mathrm{cr2}$ and $H_\mathrm{cr1}$.
The simulated domain contains four DWs as shown in Fig.~\ref{fig:T3}(a).
For three of the four DWs, the Bloch component is aligned along the field, while for the other DW it is opposite to the field, thus leading to the formation of a HBL.
The simulated magnetization curves for decreasing the field to zero and then increasing it back to $180$\,mT for samples of different thickness are shown in Fig.~\ref{fig:T3}(b).
Within a certain ML thickness regime, the curves show a thickness-dependent opening, which can be defined with two critical fields $H^*_\mathrm{cr2}$ and $H_\mathrm{cr2}$, corresponding to the switching field of the descending and, respectively, ascending branch. Note that the critical field $H_\mathrm{cr2}$ is coinciding with the previous definition (indicated in Fig.~\ref{fig:T1}(g)). 
Figure~\ref{fig:T3}(c) shows the dependence of the critical fields $H^*_\mathrm{cr2}$ and $H_\mathrm{cr2}$ on the ML thickness.
According to the simulation result, below $X=50$, the formation of the HBL represents a fully reversible process, while for thicker MLs the hysteretic behavior emerges with an increasingly more pronounced hysteresis.
For very thick MLs the simulations predict that DWs with HBL can be stable even at zero field, such that a reverse magnetic field is required for unwinding.
We confirmed the theoretical prediction for the opening of minor loops near $H_\mathrm{cr2}$ by experimental measurements, see the insets in Fig.~\ref{fig:Fig3}(a)-(c).
It is seen that the magnetization curve is indeed hysteresis free for $X=48$ and that the opening increases with the film thickness. 
From above we conclude that the theoretical picture based on the appearance of HBL is correct.

Still, we would like to note that full quantitative agreement between simulation and experiment is not attained, as the simulated remanent magnetizations are about half the experimental values.
We identified a few reasons for this discrepancy.
First, we note that the period of experimentally aligned stripe domains via in-plane demagnetization might be lower than the actual ground state equilibrium period.
In particular, the period of a labyrinth domain pattern obtained after out-of-plane demagnetization typically shows a periodicity approximately up to $50\,\%$ larger than for aligned stripe domains~\cite{6,Kozlov}. 
It has been suggested recently, that the labyrinth pattern in magnetic MLs is closer to the ground state equilibrium period than the aligned stripe domain pattern, because the formation of the latter via in-plane demagnetization might be more susceptible to defects~\cite{Kozlov}.
As follows from Fig.~\ref{fig:T1}(f), a higher periodicity corresponds to higher exchange stiffness, and as a result larger width of DWs, such that a higher in-plane magnetization might be obtained in the simulations.  
Moreover, the anisotropy value used for the simulation was experimentally determined for a low-$X$-sample, however, the anisotropy might be different for thicker samples. 
Additionally, there might be a perpendicular anisotropy gradient present as the average out-of-plane lattice spacing relaxes from an initially strained state (due to the Pt seed layer) to its equilibrium value, similar as has been observed in recent studies~\cite{Fallarino} for MLs with $X>20$.

\section{Conclusions}
We study the domain wall magnetization switching in aligned stripe domain states in [(Co(0.44 nm)/Pt(0.7 nm)]$_X$ ($X = 48$, 100, 150) multilayers with perpendicular magnetic anisotropy. While no remanent in-plane moment is present in the in-plane AC demagnetized state, we are able to induce a net remanent moment after the application of an in-plane magnetic field parallel to the stripes. We demonstrate that this in-plane moment originates from the polarization of the Bloch-type DWs.
The DW polarization as well as its full reversal have no influence on the stripe domain alignment and morphology itself.
This permits the study of the DW magnetization reversal by recording the remanent magnetization after the application of different magnetic field amplitudes ($H_\mathrm{m}$) along the stripe domain direction.
The reversal process shows a double step behavior with the DW activation field ($H_\mathrm{cr2}$) and DW saturation field ($H_\mathrm{cr1}$) as listed in Table I. 
The reversal mechanism is attributed to a DW magnetization twist in localized regions, similar to HBLs introduced by Malozemoff and Slonczewski for garnet films~\cite{Malozemoff_79}.
At $H_\mathrm{cr2}$ the HBLs emerge in DWs polarized opposite to the applied field. Some percentage of these DWs then undergo a possibly defect-induced switching, yielding a net remanent moment when reducing the field back to 0. At $H_\mathrm{cr1}$, finally, all HBLs collapse, resulting in full polarization of the DWs and maximal remanent moment. This transition, however, is smoothed in the hysteresis loops due to film inhomogeneities.

Our studies provide the opportunity for experimental estimation of the effective DW width using only magnetometry and conventional magnetic force microscopy methods.
For $Q \gtrsim 1$ the effective DW width is practically independent of the ML thickness.
Furthermore, the DW magnetization switching at small external fields, opens a pathway to control the DW magnetization with high precision independent of the domain pattern. Possibly, the DW magnetization switching might be also achieved using a spin polarized electrical current. This, however, requires further investigations.

Our results are relevant for studies of recently discovered magnetic multilayers with interfacial Dzyaloshinskii-Moriya interaction (iDMI)~\cite{2,34,35,36}. It is argued that the iDMI term supports the N\'{e}el component of the DW to be preferred over the Bloch component with some intermediate or hybrid states, depending on the iDMI strength~\cite{37,38}. For high-Q materials the in-plane remanent moment, measured parallel to the DW direction, should be characteristic for the portion of the Bloch component. The N\'{e}el-type modulations inside the DW in ML with iDMI should provide zero remanent magnetization in the plane of the film.

Lastly, the control of the DW polarization in magnetic stripe domain configurations can be applied in magnonics, where stripe DWs are used as a medium for spin wave propagation, resembling a self-organized magnonic crystal~\cite{3,3a,4,4a,13}.
For example, an alternating Bloch wall polarization in a stripe domain structure leads to stronger spin wave mode separation as compared to Bloch walls with the same polarization ~\cite{3a}. 
Furthermore, inversion of the DWs magnetization direction can attenuate the spin wave amplitudes due to the non-reciprocity of spin wave frequency dispersion~\cite{4a}, acting as a controller for spin wave diodes~\cite{3a,4a}.

\section{Acknowledgments}

The authors are grateful to Thomas Naumann and Jakob Heinze for experimental and technical support. 
NSK, BR and SS gratefully acknowledge financial support through the Priority Program SPP2137 "Skyrmionics" of the German Research Foundation (DFG) within projects KI 2078/1-1, RE-1164/6-1 and LU-2261/2-1.

\end{document}